\documentclass[12pt]{article}
\usepackage[left=3cm,top=3cm,right=3cm,bottom=2cm]{geometry}
\usepackage{graphicx}
\usepackage{amssymb,amsmath}

\newcommand{\upi}{\pi}

\newcommand{\bea}{\begin{eqnarray}}
\newcommand{\eea}{\end{eqnarray}}
\newcommand{\citep}[1]{\cite{#1}}
\author{Robert W. Johnson \\
\small Alphawave Research\\[-0.8ex]
\small Atlanta, GA, USA\\
\small \texttt{robjohnson@alphawaveresearch.com}\\}
\title{}
\date{\today}

\newcommand{\beq}{\begin{equation}}
\newcommand{\eeq}{\end{equation}}

\newcommand{\epsi}{\epsilon}

\newcommand{\Cth}{\cos \theta}
\newcommand{\Sth}{\sin \theta}

\newcommand{\xhat}{\hat{x}}
\newcommand{\yhat}{\hat{y}}
\newcommand{\rhat}{\hat{r}}
\newcommand{\zhat}{\hat{z}}
\newcommand{\thetahat}{\hat{\theta}}
\newcommand{\phihat}{\hat{\phi}}
\newcommand{\perphat}{\hat{\perp}}
\newcommand{\parahat}{\hat{\parallel}}
\newcommand{\para}{\parallel}
\newcommand{\wt}[1]{\widetilde{#1}}
\newcommand{\mbf}[1]{\mathbf{#1}}

\newcommand{\Bpol}{B_\theta}
\newcommand{\Btor}{B_\phi}
\newcommand{\Vpol}{V_\theta}
\newcommand{\Vtor}{V_\phi}
\newcommand{\Om}{\dfrac{V_\phi}{R}}
\newcommand{\del}{\nabla}
\newcommand{\divr}{\nabla \cdot}

\newcommand{\ddr}[1]{\dfrac{\partial\, {#1}}{\partial r}}
\newcommand{\ddp}[1]{\dfrac{\partial\, {#1}}{\partial \theta}}
\newcommand{\ddrp}[1]{\dfrac{\partial\, {#1}}{r \partial \theta}}
\newcommand{\dddrr}[1]{\dfrac{\partial^2\, {#1}}{\partial r^2}}

\newcommand{\oover}[1]{\dfrac{1}{#1}}
\newcommand{\dsub}[1]{\partial_{#1}}
\newcommand{\bp}{b_\theta}
\newcommand{\bt}{b_\phi}

\begin{document}

\maketitle

\begin{abstract}
The model by Braginskii for the viscous stress tensor is used to determine the shear and gyroviscous forces acting within a toroidally confined plasma.  Comparison is made to a previous evaluation which contains an inconsistent treatment of the radial derivative and neglects the effect of the pitch angle.  Parallel viscosity contributes a radial shear viscous force which may develop for sufficient vertical asymmetry to the ion velocity profile.  An evaluation is performed of this radial viscous force for a tokamak near equilibrium which indicates qualitative agreement between theory and measurement for impure plasma discharges with strong toroidal flow.
\end{abstract}


\section{Introduction}
The evaluation of the Braginskii model of plasma viscosity~\citep{brag-1965} for toroidal confinement is hampered by the complicated geometry.  Previous attempts~\citep{frc-pop-2006,stacandsig-1985,shaingetal-1985} have not treated the radial derivative in a consistent manner, nor have they fully taken into account the effect of the magnetic field pitch angle on the viscous stress tensor.  As the determination of this quantity is of intrinsic interest to the plasma theorist as well as to fusion engineers, here we reevaluate the shear and gyroviscous forces for a toroidal plasma.

This paper focuses on the viscosity model derived by Braginskii for a large flow velocity $V \sim v_{th}$ and assumes a ratio of gyrofrequency to self-collision frequency $\omega / \nu \equiv \omega \tau \gg 1$.  We note the recent derivations by Mikhailovskii and Tsypin~\citep{mikntyspin-1051}, Catto and Simakov~\citep{simakov-4744,catto-1190,catto-012501}, and Ramos~\citep{ramos-112301} consider an ordering with small flow velocity $V \ll v_{th}$ which introduces terms dependent on the heat flux.  We will neglect such effects as well as any arising from trapped particles~\citep{shaingetal-1985,hinton-3082,wong-112505,connor-29919} as a first approximation to the extension of the classical Braginskii viscosity to toroidal geometry.  In situations where the assumptions of applicability are not met, these additional terms may be necessary.

After working through in detail a prevailing derivation of the toroidal gyroviscous torque, we consider an alternative approach based on the geometric nature of tensor quantities.  For both models we will use the concentric, circular flux surface approximation; however, the treatments of the radial derivative will be distinct.  Restricting consideration to flux surfaces with constant density yet allowing for vertical asymmetry to the ion toroidal velocity profile yields a radial shear viscous force which should be accounted for when interpreting experimental measurements~\citep{solomonetal-pop-2006,stacey-cpp06}.  An evaluation from data collected at DIII-D~\citep{diiid-2002} indicates qualitative agreement between theory and measurement for impure plasma discharges with significant toroidal rotation.

\section{Stacey-Sigmar model for gyroviscous torque}
We begin by considering in detail the derivation of the gyroviscous torque as given by Stacey and Sigmar~\citep{stacandsig-1985}.  Using a concentric, circular toroidal flux surface geometry $(r,\theta,\phi)$ with major radius $R_0$ allows one to write $R = R_0 + r \Cth \equiv R_0 (1 + \epsi \Cth)$ and $\mbf{B} = \mbf{B}^0 / (1 + \epsi \Cth)$ with $\divr \mbf{B} = 0$ to set $B_r = 0$.  Derivatives $\partial / \partial x$ may be written $\dsub{x}$, and we assume toroidal symmetry $\dsub{\phi} \rightarrow 0$.  Considering a single ion species, the density on the flux surface is expanded about a radially dependent constant as $n = n^0(r) [1 + \epsi (\wt{n}^c \Cth + \wt{n}^s \Sth)]$, with similar expressions for the components of the fluid velocity $\mbf{V}$, and the radial dependence $\dsub{r} \equiv R_0^{-1} \dsub{\epsi}$ of the poloidal variations are neglected when taking the radial derivative of an expanded quantity, $\dsub{r} n \approx (n/n^0) \dsub{r} n^0$, despite the explicit appearance of the factor $\epsi$ as well as any radial dependence of the poloidal coefficients $\wt{n}^{c,s}(r)$.  The poloidal derivative is taken explicitly, $\dsub{\theta} n = n^0 \epsi (\wt{n}^s \Cth - \wt{n}^c \Sth)$.  In this section, the thermal energy $k_B T \rightarrow T = m v_{th}^2 / 2$ assumes no poloidal variation so that the pressure takes on the dependence of the density, $p = n T = p^0 (n/n^0)$, for consistency of comparison.  One assumes here that both the ions and the electrons are isothermal.

For an equilibrium density $\dsub{t} n \rightarrow 0$ the continuity equation reads $\dot{n} \equiv d n / d t = \divr n \mbf{V} = (\dsub{r} r R n V_r + \dsub{\theta} R n \Vpol) / r R$, where $\dot{n}$ is the particle source rate.  The unity moment flux surface average \beq
\left< A \right>_U \equiv \oover{2 \upi} \oint_0^{2\upi} d\theta \left( 1+\epsi\Cth \right) A(\theta) \;, \eeq 
yields the differential equation $\dot{n}^0 = n^0 V_r^0 / r + n^0 \dsub{r} V_r^0 + V_r^0 \dsub{r} n^0$ which has nontrivial solution for the radial velocity only with nonzero source rate; for $\dot{n} \rightarrow 0$, $V_r \rightarrow 0$.  Then the cosine \beq
\left< A \right>_C \equiv \oover{2 \upi} \oint_0^{2\upi} d\theta \Cth \left( 1+\epsi\Cth \right) A(\theta) \eeq 
and sine $\left< A \right>_S$ moments relate the density variations to those of the poloidal velocity, $\wt{\Vpol}^c = - \wt{n}^c - 1$ and $\wt{\Vpol}^s = - \wt{n}^s$.  The flux surface averaged gyroviscous torque is derived from the rate of shear tensor under the assumption that $\left| \Btor / B \right| \approx 1$ and $\left| \Bpol / B \right| \approx 0$ as \beq
\left< R \phihat \cdot \divr \mbf{\Pi}^{34} \right> \simeq - \left< \oover{r R} \ddr{} \eta_4 R^3 \ddp{} \Om \right> \;, \eeq
where $\eta_4 \simeq n T m / z e B$ is the gyroviscous coefficient.  We split the integrand into five terms \bea 
I_1 & = & (-r R)^{-1} \left( \ddr{p^0} \right) \dfrac{n}{n^0} \dfrac{\eta_4}{p} R^3 \ddp{} \Om \;,\\
I_2 & = & (-r R)^{-1} \eta_4 \dfrac{B}{B^0} \left( \ddr{} \dfrac{B^0}{B} R^3 \right) \ddp{} \Om \;, \\
I_3 & = & (-r R)^{-1} \eta_4 R^3 B^0 \left( -B^0 \right)^{-2} \left( \ddr{B^0} \right) \ddp{} \Om \;,\\
I_4 & = & (-r R)^{-1} \eta_4 R^3 \ddp{} \left( \ddr{\Vtor^0} \right) \dfrac{\Vtor}{\Vtor^0 R} \;,\\
I_5 & = & (-r R)^{-1} \eta_4 R^3 \ddp{} \Vtor \ddr{} \oover{R} \;,  \eea 
collecting their units into the common factor $I_U \equiv n^0 T m \Vtor^0 / 2 z e B^0 R_0 = \eta_4^0 \Omega^0 / 2$, and evaluate their leading contributions thusly, \bea 
\left< I_1 / I_U \right> & \simeq & r L_p^{-1} \left[ \wt{V}_\phi^s \left( 4 + \wt{n}^c \right) + \wt{n}^s \left( 1 - \wt{V}_\phi^c \right) \right] \;, \\
\left< I_2 / I_U \right> & \simeq & -4 \wt{V}_\phi^s  \;, \\
\left< I_3 / I_U \right> & \simeq & -r L_B^{-1} \left[ \wt{V}_\phi^s \left( 4 + \wt{n}^c \right) + \wt{n}^s \left( 1 - \wt{V}_\phi^c \right) \right] \;, \\
\left< I_4 / I_U \right> & \simeq & r L_{\Vtor}^{-1} \left[ \wt{V}_\phi^s \left( 4 + \wt{n}^c \right) + \wt{n}^s \left( 1 - \wt{V}_\phi^c \right) \right] \;, \\
\left< I_5 / I_U \right> & \simeq & - \wt{n}^s  \;,  \eea 
where $L_X^{-1} \equiv - \dsub{r} \ln X^0$ is the inverse radial gradient scale length for $X$, giving a net gyroviscous torque of \beq \label{eqn-gyrotorq}
\left< R \phihat \cdot \divr \mbf{\Pi}^{34} \right> \simeq \dfrac{- \eta_4^0 \Omega^0}{2} \left\lbrace 4 \wt{V}_\phi^s + \wt{n}^s - r \left( \sum L_X^{-1} \right) \left[ \wt{V}_\phi^s \left( 4 + \wt{n}^c \right) + \wt{n}^s \left( 1 - \wt{V}_\phi^c \right) \right] \right\rbrace \;, \eeq
where $\sum L_X^{-1} \equiv L_p^{-1} + L_{\Vtor}^{-1} - L_B^{-1}$.  The Stacey-Sigmar model is seen to correspond to the contribution from $I_1 + I_4$.  The evaluation of $I_2$ and $I_5$ arises from considering the poloidal dependence of $R$ and $B$ as geometrically induced rather than treating them as expanded quantities, and the term $I_3$ represents the change in the gyroviscous coefficient dependent upon the gyrofrequency, as $\dsub{r} B^0 = (\Bpol^0 / B^0) \dsub{r} \Bpol^0$ when $\dsub{r} \Btor^0 = 0$.  Interestingly, the leading terms in (\ref{eqn-gyrotorq}) are of opposite sign, hence would tend to cancel, those of the Stacey-Sigmar model.  The derivation for the poloidal shear viscosity~\citep{frc-pop-2006,shaingetal-1985} proceeds similarly with the same caveats and pitfalls.

\section{Viscous force for toroidal confinement}
Next we consider an alternate derivation making use of the transformation properties of the viscosity tensor.  We also will treat the radial derivative consistently throughout.  While our expansion looks the same, $n = n^0(r) [1 + \epsi (\wt{n}^c \Cth + \wt{n}^s \Sth)]$, here we specify the poloidal coefficients to have no radial dependence, $\dsub{r} \wt{n}^{c,s} = 0$, and the radial derivative is explicitly $\dsub{r} n = (n/n^0) \dsub{r} n^0 + n^0 (n/n^0 - 1) / r$.  The unity moment of the continuity equation remains the same, and remembering that physically it is the dynamic equation for the density, we use its cosine and sine moments to write $\wt{n}^c = - \wt{V}_\theta^c -1$ and $\wt{n}^s = - \wt{V}_\theta^s$ when $V_r \rightarrow 0$.  That we have used its unity moment to determine $V_r^0$ needs to be recalled when addressing the radial component of the equation of motion.  In the presence of a radial flow, the density coefficients acquire a correction factor $\wt{n}^{c,s} \rightarrow \wt{n}^{c,s} / (1 + 2 r \dsub{r} \ln n^0)$ which would introduce an $r$ dependence unless $n^0 \sim r$.  The poloidal dependence of the ion temperature is addressed through the equilibrium equation of state $\mbf{V} \cdot \del p + \gamma p \divr \mbf{V} = 0$ for adiabatic index $\gamma$, with solution to $\mathcal{O}(\epsi)$ of $\wt{T}_i^{c,s} = \wt{\gamma}_i \wt{n}_i^{c,s}$ for $\wt{\gamma}_i \equiv \gamma_i - 1$.

\begin{figure}[]
\includegraphics[scale=.4]{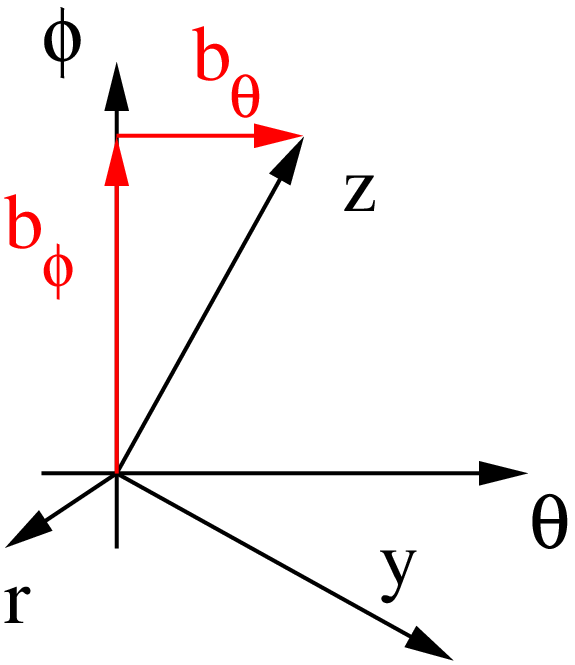}%
\caption{\label{fig1} The transformation of vectors and tensors between orthonormal coordinate axes is given by the rotation matrix, which may be found from the components of one set of unit vectors along the other set of coordinate axes.  Here, $\yhat = \bt \thetahat - \bp \phihat$ and $\zhat = \bp \thetahat + \bt \phihat$. }
\end{figure}

Moving on, the rate of shear tensor in general is written as \bea 
W_{\alpha \beta} & \equiv & \hat{\alpha} \cdot \del \mbf{V} \cdot \hat{\beta} + \hat{\beta} \cdot \del \mbf{V} \cdot \hat{\alpha} - \dfrac{2}{3} \delta_{\alpha \beta} \divr \mbf{V} \;, \\
& = & \dsub{\alpha} V_\beta + \sum_\gamma \Gamma^\alpha_{\beta \gamma} V_\gamma + \dsub{\beta} V_\alpha + \sum_\gamma \Gamma^\beta_{\alpha \gamma} V_\gamma - \dfrac{2}{3} \delta_{\alpha \beta} \divr \mbf{V} \;,
 \eea and reduces for toroidally symmetric $(r,\theta,\phi)$ coordinates with non-vanishing Christoffel symbols $\Gamma^\phi_{\phi r} = -\Gamma^\phi_{r \phi} = \Cth / R$, $\Gamma^\phi_{\phi \theta} = -\Gamma^\phi_{\theta \phi} = -\Sth / R$, and $\Gamma^\theta_{\theta r} = -\Gamma^\theta_{r \theta} = 1/r$ and vanishing radial flow to \beq
\mbf{W}' = \left[ \begin{array}{lll} 0 & \ddr{\Vpol} - \dfrac{\Vpol}{r} & \ddr{\Vtor} - \dfrac{\Cth}{R} \Vtor \\ W_{r \theta} & 2 \ddrp{\Vpol} & \ddrp{\Vtor} + \dfrac{\Sth}{R} \Vtor \\ W_{r \phi} & W_{\theta \phi} & - 2 \dfrac{\Sth}{R} \Vpol \end{array} \right] - \dfrac{2}{3} \left( \divr \mbf{V} \right) \mbf{I} \;, \eeq 
which is a geometric object independent of one's choice of coordinate system.  Identifying $(\rhat,\thetahat,\phihat)$ as the primed coordinate system, the transformation to an unprimed coordinate system $(\xhat,\yhat,\zhat)$ is given by $\mbf{W} = \mbf{G}^T \mbf{W}' \mbf{G}$, where $\mbf{G} \in \mathrm{SO}(3)$ is a general parity preserving rotation matrix in ${\Re}^3$.  Braginskii's theory is derived for coordinate axes orientated to the magnetic field, thus for $(\xhat,\yhat,\zhat) \equiv (\rhat,\perphat,\parahat)$ where $\parahat \equiv \mbf{b} \equiv \mbf{B}^0/B^0 = (0,\bp,\bt)$ in $(\rhat,\thetahat,\phihat)$, the rotation matrix is \beq
\mbf{G} = \left[ \begin{array}{ccc} 1 & 0 & 0 \\ 0 & \bt & \bp \\ 0 & -\bp & \bt \end{array} \right] \;, \eeq 
as given in Fig.~\ref{fig1}.  Our strategy is to take $\mbf{W}' \rightarrow \mbf{W} \rightarrow \mbf{\Pi} \rightarrow \mbf{\Pi}'$, where $\mbf{\Pi}' = \mbf{G} \mbf{\Pi} \mbf{G}^T$.  The viscosity coefficients are in the ordering $\eta_0 \gg \eta_4 \gg \eta_2$ for a strong magnetic field $\omega \tau \gg 1$, and we will neglect the perpendicular contribution $\eta_1, \eta_2 \rightarrow 0$.  The shear viscosity reduces to $\mbf{\Pi}_S = - \eta_0 \mbf{W}_0$, and the gyroviscosity to $\mbf{\Pi}_G = \eta_4 (\mbf{W}_3/2 + \mbf{W}_4)$, where $\eta_0 \simeq n T \tau$ and $\eta_4 \simeq n T m / z e B$ such that $\eta_0 / \eta_0^0 = (T/T^0)^{5/2}$ and $\eta_4 / \eta_4^0 = (n/n^0)(T/T^0)(R/R_0)$.  Explicitly, the shear component is \beq
\mbf{\Pi}_S = - \eta_0 \left[ \begin{array}{lll} (W_{r r} + W_{\perp \perp})/2 & 0 & 0 \\ 0 & (W_{r r} + W_{\perp \perp})/2 & 0 \\ 0 & 0 & W_{\para \para} \end{array} \right] \;, 
\eeq while the gyroviscous component is \beq
\mbf{\Pi}_G = \eta_4 \left[ \begin{array}{lll} - W_{r \perp}/2  & (W_{r r} - W_{\perp \perp})/4 & -W_{\perp \para} \\ (W_{r r} - W_{\perp \perp})/4 & W_{r \perp}/2 & W_{r \para} \\ -W_{\para \perp} & W_{\para r} & 0 \end{array} \right] \;, \eeq 
 and the transformation is given by \beq
\mbf{\Pi}' = \left[ \begin{array}{lll}
 \Pi_{r r} & \bt \Pi_{r \perp}+\bp \Pi_{r \para} & -\bp \Pi_{r \perp}+\bt \Pi_{r \para} \\ 
\Pi_{r \theta} & 2 \bt \bp \Pi_{\perp \para}+\bt^2 \Pi_{\perp \perp}+\bp^2 \Pi_{\para \para} & (\bt^2-\bp^2) \Pi_{\perp \para}-\bt \bp (\Pi_{\perp \perp}-\Pi_{\para \para}) \\ 
\Pi_{r \phi} & \Pi_{\theta \phi} & -2 \bt \bp \Pi_{\perp \para}+\bp^2 \Pi_{\perp \perp}+\bt^2 \Pi_{\para \para}
\end{array} \right] \;. \eeq 

In our evaluation of the viscosity, all derivatives are taken explicitly, neglecting only the poloidal dependence of the Coulomb logarithm.  The components of the viscous force are given by $(\divr \mbf{\Pi})_\alpha = \sum_\beta [ \dsub{\beta} \Pi_{\beta \alpha} + \sum_\gamma ( \Gamma^\beta_{\alpha \gamma} \Pi_{\beta \gamma} + \Gamma^\beta_{\beta \gamma} \Pi_{\gamma \alpha} ) ]$, and we write the toroidal viscous torque \beq
 R \phihat \cdot \divr \mbf{\Pi}' = \left( \dfrac{\Cth}{R} + \oover{r} + \ddr{} \right) R \Pi_{r \phi} - \left( \dfrac{\Sth}{R} - \ddrp{}  \right) R \Pi_{\theta \phi} \;, \eeq 
noting there is no $\eta_0$ contribution to $\Pi_{r \phi}$.  Evaluating its leading components, we find that the shear viscous torque vanishes, $\langle R \phihat \cdot \divr \mbf{\Pi}'_S \rangle \simeq 0$, and the gyroviscous torque is \beq
\begin{array}{cl} \multicolumn{2}{l}{\left< R \phihat \cdot \divr \mbf{\Pi}'_G \right> \simeq} \\
  & \dfrac{\eta_4^0}{R_0} \Vtor^0 \bt \left(\bt^2+2 \bt^2 \bp^2-\bp^4\right) \left\lbrace\left[\left(\wt{V}_\theta^c+1\right) \wt{V}_\phi^s-\left(\wt{V}_\phi^c-1\right) \wt{V}_\theta^s\right] \gamma -4 \wt{V}_\phi^s\right\rbrace \\
+ & \dfrac{\eta_4^0}{R_0} \Vpol^0 \bp \wt{V}_\theta^s \left[ \dfrac{1}{4} \left(28 \bt^2 \bp^2+7 \bt^2+4 \bp^4-\bp^2\right) \wt{\gamma} - \dfrac{1}{6} \left(4 \bp^2+1\right) \left(\bp^2-1+10 \bt^2\right) \right] \;, \end{array} 
\eeq
whereupon restricting consideration to flux surfaces with constant density, $\wt{V}_\theta^s=0$ and $\wt{V}_\theta^c=-1$, denoted by the symbol $\Rightarrow$, yields a non-vanishing torque \beq
\langle R \phihat \cdot \divr \mbf{\Pi}'_G \rangle \Rightarrow  -4 (\eta_4^0/R_0) \Vtor^0 \bt (\bt^2+2 \bt^2 \bp^2-bp^4) \wt{V}_\phi^s \;.
\eeq
 The poloidal viscous force is \beq
\thetahat \cdot \divr \mbf{\Pi}' = \oover{R} \left[ \left( \dfrac{2}{r} + \ddr{} \right) R \Pi_{r \theta} + \ddrp{} R \Pi_{\theta \theta} \right] + \dfrac{\Sth}{R} \Pi_{\phi \phi} \;, \eeq
where $\Pi_{r \theta}$ has only an $\eta_4$ component, and its shear contribution evaluates to \beq
\begin{array}{cl} \multicolumn{2}{l}{\left< \thetahat \cdot \divr \mbf{\Pi}'_S \right> \simeq} \\
  & \dfrac{\eta_0^0}{R_0^2}\Vtor^0 \bt \bp \dfrac{1}{2} \left(2 \bt^2-\bp^2\right) \left(\wt{V}_\phi^c-1\right) \\
+ & \dfrac{\eta_0^0}{R_0^2}\Vpol^0 \left[ \dfrac{1}{2} \left(2 \bt^2-\bp^2\right) \left(\bt^2-\bp^2\right) - \dfrac{1}{6} \left(\bp^4-6 \bp^2 \bt^2+\bp^2+2 \bt^4\right) \left(\wt{V}_\theta^c+1\right) \right] \;, \end{array}
\eeq which reduces to \beq
\left< \thetahat \cdot \divr \mbf{\Pi}'_S \right> \Rightarrow (\eta_0^0 / R_0^2) (\bt^2-\bp^2/2) \left[\Vtor^0 \bt \bp (\wt{V}_\phi^c-1) + \Vpol^0(\bt^2-\bp^2) \right] \;.
\eeq Its gyroviscous component yields \beq
\begin{array}{cl} \multicolumn{2}{l}{\left< \thetahat \cdot \divr \mbf{\Pi}'_G \right> \simeq} \\
  & \dfrac{\eta_4^0}{R_0^2}\Vtor^0 \bp \left( \dfrac{1}{2} \left[\left(-1-2 \bp^2\right) \bt^2+4 \bp^4\right] \left[\left(\wt{V}_\phi^c-1\right) \wt{V}_\theta^s \gamma -\left(\wt{V}_\theta^c+1\right) \wt{V}_\phi^s \wt{\gamma}\right] \right. \\
  & \multicolumn{1}{r}{\left. + \dfrac{1}{4} \left\lbrace\left[\left(4 \bp^2+2\right) \bt^2-8 \bp^4\right] \left(\wt{V}_\theta^c+1\right) -2 \left(6 \bp^2+5\right) \bt^2+24 \bp^4-\bp^2\right\rbrace \wt{V}_\phi^s \right)} \\
+ & \dfrac{\eta_4^0}{R_0^2}\Vpol^0 \bt \wt{V}_\theta^s \left\lbrace \dfrac{1}{4} \left[\left(4 \bp^2+3\right) \bt^2+28 \bp^4+4 \bp^2\right] \wt{\gamma} \right. \\
  & \multicolumn{1}{r}{\left. - \dfrac{1}{6} \left[\left(4 \bp^2+3\right) \bt^2+22 \bp^4+11 \bp^2+2\right] \right\rbrace \;, } \end{array}
 \eeq and it reduces to \beq
\left< \thetahat \cdot \divr \mbf{\Pi}'_G \right> \Rightarrow - \dfrac{\eta_4^0}{R_0^2}\Vtor^0 \bp \wt{V}_\phi^s \dfrac{1}{4} \left[2 \left(6 \bp^2+5\right) \bt^2+24 \bp^4-\bp^2\right] \;. 
\eeq The radial viscous force is \beq
\rhat \cdot \divr \mbf{\Pi}' = \oover{r R} \left[ \left( 1 + r \ddr{} \right) R \Pi_{r r} + \ddp{} R \Pi_{r \theta} \right] - \oover{r} \Pi_{\theta \theta} - \dfrac{\Cth}{R} \Pi_{\phi \phi} \;, 
\eeq and its shear component is \beq \label{eqn-4parama} 
\begin{array}{cl} \multicolumn{2}{l}{\left< \rhat \cdot \divr \mbf{\Pi}'_S \right> \simeq} \\
  & \dfrac{\eta_0^0}{R_0^2} \Vtor^0 \bt \bp \left( \left(1+ \dfrac{7}{2} \bp^2 \right) \wt{V}_\phi^s \right.\\
  & \multicolumn{1}{r}{\left. + \dfrac{5}{4} \left\lbrace \left(4 \bp^2-\bt^2+1\right) \left[\left(\wt{V}_\phi^c-1\right) \wt{V}_\theta^s -\left(\wt{V}_\theta^c+1\right) \wt{V}_\phi^s\right] \wt{\gamma} \right\rbrace \right)} \\
+ & \dfrac{\eta_0^0}{R_0^2}\Vpol^0 \wt{V}_\theta^s \left\lbrace \dfrac{5}{4} \left[-\bt^4+\left(2+5 \bp^2\right) \bt^2-4 \bp^4\right] \wt{\gamma} \right. \\
  & \multicolumn{1}{r}{\left. + \dfrac{1}{6} \left[11 \bp^4-\left(1+9 \bt^2\right) \bp^2+\bt^2 \left(\bt^2-6\right)\right] \right\rbrace \;,} \end{array} 
 \eeq
which for constant flux surface density reduces to \beq \label{eqn-radshear}
\left< \rhat \cdot \divr \mbf{\Pi}'_S \right> \Rightarrow \dfrac{\eta_0^0}{R_0^2} \Vtor^0 \bt \bp \left(1+ \dfrac{7}{2} \bp^2 \right) \wt{V}_\phi^s \;,
 \eeq
 while its gyroviscous component is \beq
\begin{array}{cl} \multicolumn{2}{l}{\left< \rhat \cdot \divr \mbf{\Pi}'_G \right> \simeq} \\
  & \dfrac{\eta_4^0}{R_0^2} \dfrac{\Vtor^0}{4} \bp \left\lbrace \left(3 \bt^2-2 \bp^2-1\right) \left[\left(\wt{V}_\phi^c-1\right) \left(\wt{V}_\theta^c+1\right)+\wt{V}_\phi^s \wt{V}_\theta^s\right] \wt{\gamma} \right.  \\
  & \multicolumn{1}{r}{\left. + \left(3 \bt^2-2 \bp^2-1\right) \left[\left(\wt{V}_\phi^c-1\right) \wt{V}_\theta^c+\wt{V}_\phi^s \wt{V}_\theta^s\right] +\left(4 \bt^2+\bp^2\right) \left(\wt{V}_\phi^c-1\right)\right\rbrace} \\
+ & \dfrac{\eta_4^0}{R_0^2} \dfrac{\Vpol^0}{4} \bt \left\lbrace \left[\left(\bt^2+6 \bp^2+2\right) \left(\wt{V}_\theta^{c\;2}+\wt{V}_\theta^{s\;2}\right) +3 \bp^2 \left(\wt{V}_\theta^c-1\right)-\bt^2-2\right] \wt{\gamma} \right.  \\
  & \multicolumn{1}{r}{\left. - \left[\left(2 \bt^2+9 \bp^2+4\right) \wt{V}_\theta^c+\left(\bt^2+6 \bp^2+2\right)\right] \right\rbrace} \\
+ & \dfrac{\eta_4^0}{2} \left\lbrace \left[ \left(\bp \dddrr{\Vtor^0}-\bt \dddrr{\Vpol^0}+ \dfrac{\mu_0 J_\phi^0}{B^0} \ddr{\Vtor^0}\right) \right. \right.  \\
  & \multicolumn{1}{r}{\left. \left. + 2 \left(\bp \ddr{\Vtor^0}-\bt \ddr{\Vpol^0}+\bt \dfrac{\Vpol^0}{r}\right) \left(\ddr{\ln \sqrt{p^0}}-\bp \dfrac{\mu_0 J_\phi^0}{B^0}\right) \right] \right.}  \\
  & \multicolumn{1}{r}{\left. - \oover{r} \left[\bp \left(3 \bt^2-2 \bp^2\right) \ddr{\Vtor^0} -\bt \left(\bt^2+6 \bp^2\right) \left(\ddr{\Vpol^0}-\dfrac{\Vpol^0}{r}\right)\right] \right\rbrace \;,} \end{array}
 \eeq
 where we have used $\dsub{r} \Bpol^0 = \mu_0 J_\phi^0 - \Bpol^0 / r$.  According to this theory, the poloidal velocity must go as a quadratic (or greater) near the magnetic axis, $\Vpol^0 \sim r^2$ as $r \rightarrow 0$, else the radial gyroviscous force will diverge.  An order of magnitude estimate indicates that this force is on par with the radial inertial force, $\langle n m (\mbf{V} \cdot \del) \mbf{V} \rangle \simeq - n^0 m {\Vpol^0}^2 / r \sim 10 \mathrm{Nt/m^3}$.

\section{Radial force from vertical asymmetry}
Returning to the radial shear viscous force for constant density flux surfaces (\ref{eqn-radshear}), we see that a radial viscous force may develop from an interplay between the magnetic field pitch angle $\bp \neq 0$ and a vertical asymmetry to the toroidal velocity $\wt{V}_\phi^s \neq 0$.  We are not aware of this result appearing previously in the literature, as it disappears for an ordering $\bp \ll \bt$.  As radial force balance is of primary importance to plasma confinement, we are interested in how the Braginskii radial shear viscosity influences tokamak operation.

\subsection{Model equation}
Considering now the effects of a single null vertical divertor configuration as commonly found in a tokamak discharge, we expect there will exist a vertical asymmetry to the plasma, particularly to the toroidal velocity.  Thus, the radial shear viscous force (in conjunction with the inertial term) $- \langle \rhat \cdot [ n m ( \mbf{V} \cdot \del ) \mbf{V} + \divr \mbf{\Pi}_S ] \rangle$ should make a contribution to the radial equation of motion, \bea
\dfrac{n^0 m}{r} {\Vpol^0}^2 - \left\langle \rhat \cdot \divr \mbf{\Pi}_S \right\rangle & \simeq & \left< \dsub{r} p - n z e \left( \Btor \Vpol - \Bpol \Vtor \right) \right> \;, \\
 & \equiv & n^0 z e E_r^{exp} \;, \eea
where the right hand side is commonly identified as the radial electrostatic field yet actually consists of experimental measurements of density, temperature, magnetic field, and impurity velocity.  Thus, we are led to a model \beq \label{eqn-radvisc}
\dfrac{n^0 m}{r} {\Vpol^0}^2 - \dfrac{\eta_0^0}{R_0^2} \Vtor^0 \bt \bp \left(1+ \dfrac{7}{2} \bp^2 \right) \wt{V}_\phi^s \simeq \dsub{r} p^0 - n^0 z e \left( \Btor^0 \Vpol^0 - \Bpol^0 \Vtor^0 \right)
\eeq with one free parameter $\wt{V}_\phi^s$, which we can compare to data taken from a tokamak.  As $V_r^0$ has been determined from the equilibrium continuity equation, the unity moment of the radial equation of motion becomes the appropriate place to make a comparison between theory and experiment.  We mention here that, for the single-fluid model, the natural degrees of freedom associated with the conservation equations of mass, energy, and momentum are $n(\mbf{V})$, $T(n,\mbf{V})$, and $\mbf{V}(n,T)$, respectively.

Restricting the range of the parameter to the region $\vert \epsi_{max} \wt{V}_\phi^s \vert = \vert \Vtor^s \vert \leq 1$ keeps the expansion for the toroidal velocity from changing sign, where $\epsi_{max} \equiv a/R_0$ for minor radius $a$.  For a geometry with $\thetahat$ downward at the outer midplane, a positive/negative value of $\Vtor^s$ indicates an increase/decrease in the toroidal velocity at the poloidal null, with an opposite effect at the upper vertical midplane.  Minimizing $\chi^2$, the sum of squared residuals, over $\Vtor^s$ yields the parameter of best fit as well as the normalized root-mean-square residual $\widetilde{\chi} \equiv (N_r^{-1} \chi^2) ^{1/2}$ for $N_r$ radial locations.

\begin{table}
  \caption{Best fit parameters}
  \label{tab-fit}
  \begin{center}
  \begin{tabular}{ccccccc}
    \hline
shot & & 119306 & 119307 & 121433 & 121453 & 122338 \\
    \hline
$\Vtor^s$ & & 0.29 & 0.50 & 0.83 & 1.00 & 1.00 \\
$\widetilde{\chi}$ & & 15.77 & 14.11 & 16.44 & 84.50 & 49.51
  \end{tabular}
  \end{center}
\end{table}

\subsection{Data Analysis}
\begin{figure}
\includegraphics[width=\textwidth]{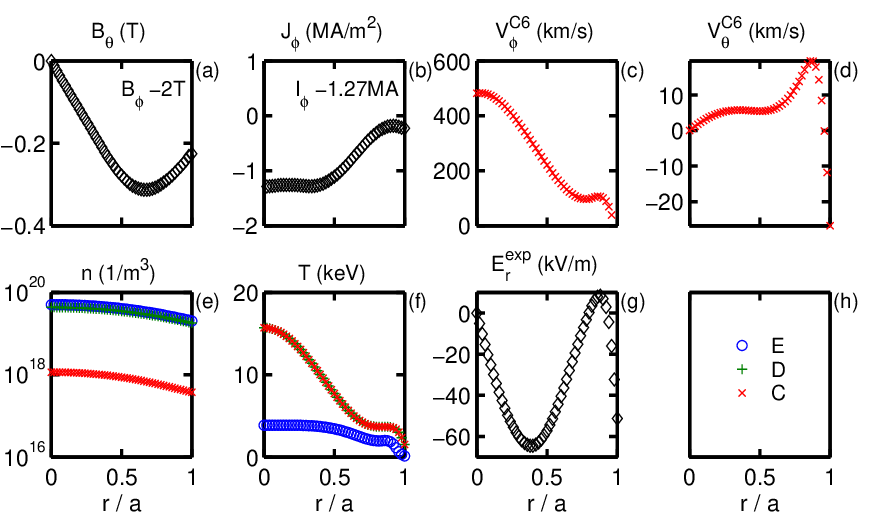}%
\caption{\label{fig2} Measured profiles for shot 119306 at time 2955ms. \vspace{1cm} }
\end{figure}

\begin{figure}[!h]
\includegraphics[width=\textwidth]{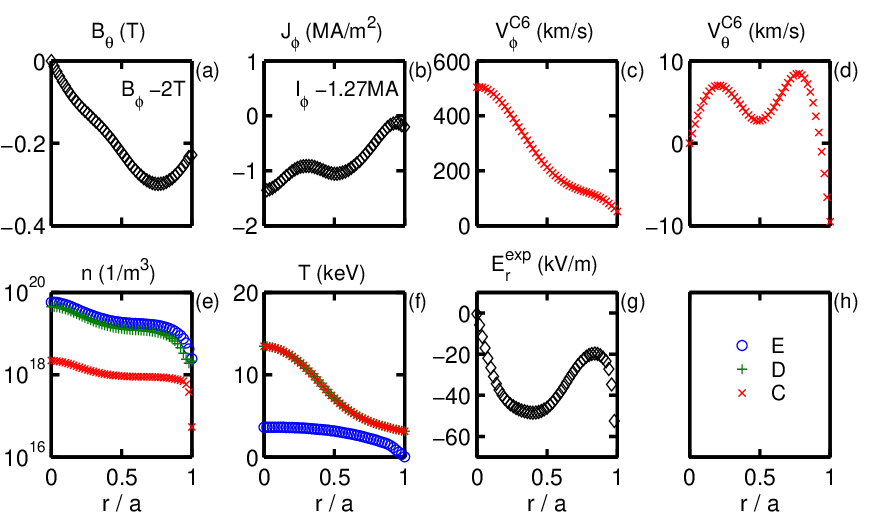}%
\caption{\label{fig3} Measured profiles for shot 119307 at time 2205ms. \vspace{1cm} }
\end{figure}

\begin{figure}
\includegraphics[width=\textwidth]{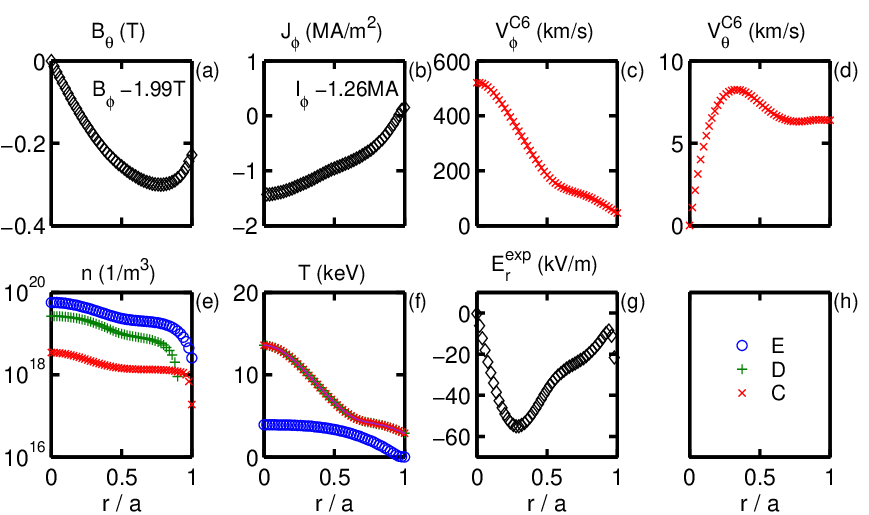}%
\caption{\label{fig4} Measured profiles for shot 121433 at time 3075ms.  He and Ne also were present. }
\end{figure}

\begin{figure}
\includegraphics[width=\textwidth]{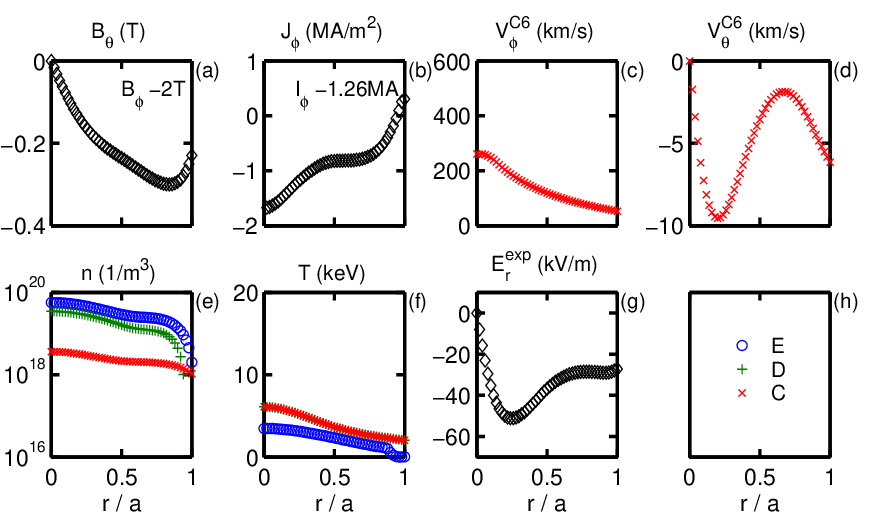}%
\caption{\label{fig5} Measured profiles for shot 121453 at time 3075ms.  He and Ne also were present.  Note the low value for the toroidal velocity (c) and ion temperature (f). }
\end{figure}

\begin{figure}
\includegraphics[width=\textwidth]{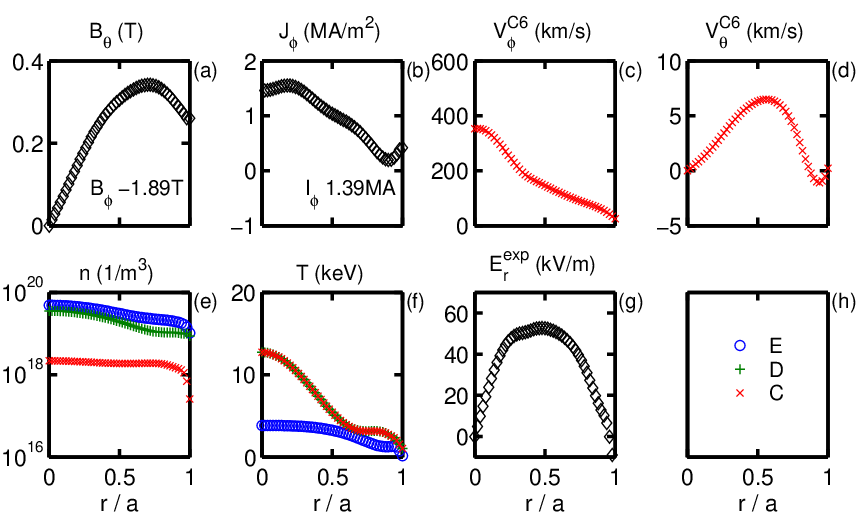}%
\caption{\label{fig6} Measured profiles for shot 122338 at time 2750ms. }
\end{figure}

To evaluate the comparison of the predicted radial viscous force with experimental measurements, we need the profiles for density, temperature, poloidal and toroidal magnetic field, and the rotation profiles for at least one species of ion.  To analyze discharges at DIII-D~\citep{diiid-2002}, we retrieve the necessary magnetic geometry and pressure profiles from EFIT~\citep{Lao:1985mw}, and the rotation profiles for $C_6$ are corrected for deuterium's energy dependent charge exchange cross section~\citep{solomonetal-pop-2006}.  Collision rates are given by the NRL Formulary~\citep{physics-nrl}.  The measurement profiles in terms of the normalized minor radius $r/a$ for our selection of shots are given in Figs.~\ref{fig2} through \ref{fig6}.  Times given are representative of the available measurement times, and the shots 121433 and 121453 have had helium and neon impurities injected.

\begin{figure}
\includegraphics[width=\textwidth]{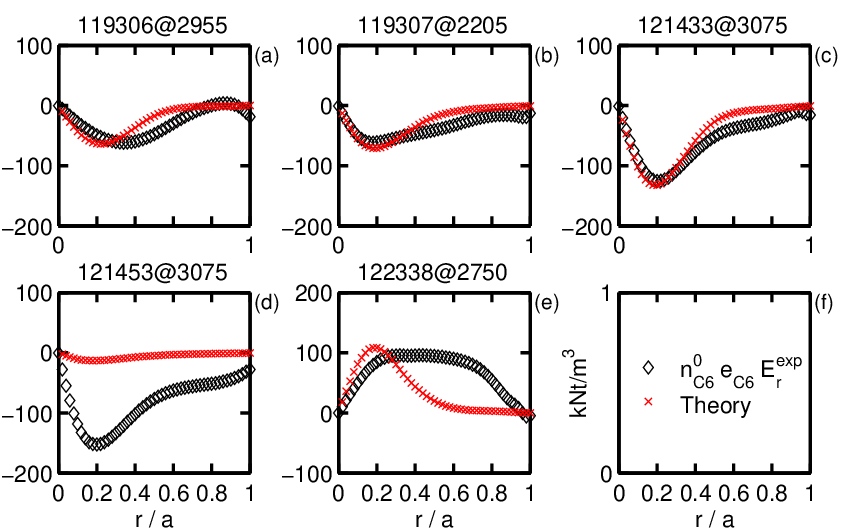}%
\caption{\label{fig7} Comparison of theory ($\times$) to measurement ($\diamondsuit$) in (\ref{eqn-radvisc}).  The parameters of best fit are in Table~\ref{tab-fit}. }
\end{figure}

\clearpage

Our results presented in Fig.~\ref{fig7} compare the measurements on the right hand side of (\ref{eqn-radvisc}) with the prediction on the left hand side, where the Theory values include the inertial term.  For three of the five shots considered, the theoretical force density can agree with the measurement for a fitted parameter $\vert \Vtor^s \vert \leq 1$ with standard error $\lesssim 10^{-5}$, as found in Table~\ref{tab-fit}, where the positive values of the parameter indicate that the plasma increases in toroidal velocity near the poloidal null and $\widetilde{\chi}$ is normalized to the units of the figure.  However, for the other two, the maximum of likelihood lies outside the prior range and the posterior is correspondingly truncated.  (A model with all 4 velocity parameters (\ref{eqn-4parama}), thus including the density and temperature poloidal variations as related above, did not noticeably improve the quality of fit, hence is discarded by Occam's principle~\citep{Sivia-1996}.)  For shot 121453, the theory cannot reach the magnitude of the measurement profile on account of the low values of the toroidal velocity and ion temperature, and for shot 122338 the shape of the profile cannot be matched by the present theory.  Were it not for these two shots, we would conclude that the Braginskii theory of viscosity is entirely consistent with the experimental measurements.  At the very least, we are forced to conclude that the combination of measurements on the right hand side of (\ref{eqn-radvisc}) need not represent an accurate evaluation of any radial electrostatic field within a tokamak.

\section{Discussion and Conclusion}
How much impact has our neglect of any heat flux had on our results?  The heat flux contribution to the generalized viscosity appears at orders consistent with the perpendicular and gyroviscosity~\citep{brag-1965} for a strong flow, though recent authors~\citep{catto-1190} suggest it contributes to the parallel viscosity as well for subsonic flow.  For a model based on an equilibrium evaluation, however, the temperature is assumed to be steady in time, hence no heat flows on the timescales of interest.  Such assumption is consistent with our assumption of constant density hence constant temperature on the flux surfaces.  The parameter $\Vtor^s$ is independent of $\Vpol^s$, thus also of $n^s$ and $T^s$.  Our results suggest that a viscous force may develop in conjunction with vertical asymmetry introduced by a divertor configuration which has not been considered previously in the literature.

Whenever one compares a theory to a measurement, one must consider whether the experiment has actualized the assumptions of applicability.  In a real tokamak, a multitude of processes may be occurring which are not considered here, and the failure of the model to account for the radial force density profile of two of the five shots may reflect these neglected phenomena.  We are intrigued by shots 121433 and 121453, which have exceedingly similar radial force density profiles yet quite different temperature and velocity profiles.  The evaluation of the Braginskii viscous force is straightforward using axes rotation, and the introduction of higher order effects does not obviously detract from the parallel contribution.  Certainly improvements to the geometry to handle flux surfaces of arbitrary cross sectional shape can be considered~\citep{stacey-082501}; however, the major discrepancy of the concentric circular flux surface approximation is in the horizontal direction, not the vertical direction driving the radial viscous force.  Nevertheless, the qualitative agreement found for the majority of the shots considered here suggests the radial shear viscous force is an effect worthy of further investigation.

The Braginskii model for plasma viscosity in a strong magnetic field has been evaluated in the concentric circular flux surface approximation for toroidal confinement retaining the parallel and gyroviscous contributions by means of the rotation matrix between coordinate axes in the lab frame and those aligned to the magnetic field.  Explicit attention is paid to the radial derivative of expanded quantities.  Restriction to flux surfaces of constant density yields a model allowing for vertical asymmetry in the impurity ion toroidal velocity which drives a radial shear viscous force with a magnitude on the order of that attributed to the radial electric field.  A prediction for the vertical asymmetry to the toroidal velocity obtains from fitting the radial shear viscous force to the experimentally measured radial force density.  An evaluation from data indicates qualitative agreement between the theoretical and measured force densities for several discharges with strong toroidal flow.





\bibliographystyle{unsrt}


\end{document}